\documentclass[prl,nofootinbib,preprintnumbers]{revtex4}

\usepackage{amsfonts}
\usepackage{mathrsfs}
\usepackage{amssymb}
\usepackage{amsmath}
\usepackage{graphicx,epsfig}

\def\lsim{\mathrel{\mathpalette\@versim<}}

\def\gsim{\mathrel{\mathpalette\@versim>}}

\begin{document}

\preprint{CERN-PH-TH/2009-247, ACT-13-09, MIFP-09-50}

\title{D-Foam Phenomenology: Dark Energy, the Velocity of Light and a Possible D-Void}

\author{John Ellis}
\affiliation{CERN, Theory Division, CH 1211 Geneva 23, Switzerland.}

\author{Nick E. Mavromatos}
 \affiliation{King's College London, Department of Physics, London WC2R 2LS, U.K.}

\author{Dimitri V. Nanopoulos}

\affiliation{George P. and Cynthia W. Mitchell Institute for
Fundamental Physics,
 Texas A$\&$M University, College Station, TX 77843, USA }

\affiliation{Astroparticle Physics Group,
Houston Advanced Research Center (HARC),
Mitchell Campus, Woodlands, TX 77381, USA}

\affiliation{Academy of Athens, Division of Natural Sciences,
 28 Panepistimiou Avenue, Athens 10679, Greece }

\begin{abstract}

In a D-brane model of space-time foam, there are contributions to the
dark energy that depend on the D-brane velocities and on
the density of D-particle defects. The latter may also reduce the
speeds of photons {\it linearly} with their energies, 
establishing a phenomenological connection with
astrophysical probes of the universality of the velocity of light.
Specifically, the cosmological dark energy density measured at the present epoch
may be linked to the apparent retardation of energetic photons propagating from nearby
AGNs. However, this nascent field of `D-foam phenomenology' may be complicated by
a dependence of the D-particle density on the cosmological epoch.
A reduced density of D-particles at redshifts $z \sim 1$ - a `D-void' - would increase the
dark energy while suppressing the vacuum refractive index, and thereby might reconcile
the AGN measurements with the relatively small retardation seen for the
energetic photons propagating from GRB 090510, as measured by the
Fermi satellite.

\end{abstract}

\maketitle


\section{Introduction to D-Phenomenology}

The most promising framework for a quantum theory of gravity is string theory,
particularly in its non-perturbative formulation known as M-theory.
This contains solitonic configurations such as D-branes~\cite{polchinski}, including
D-particle defects in space-time. One of the most challenging problems
in quantum gravity is the description of the vacuum and its properties.
At the classical level, the vacuum may be analyzed using the tools of
critical string theory. However, we have argued~\cite{emn1} that a consistent
approach to quantum fluctuations in the vacuum, the so-called `space-time
foam', needs the tools of non-critical string theory. As an example,
we have outlined an approach to this problem in which D-branes
and D-particles play an essential role~\cite{Dfoam2,Dfoam}. Within this approach, we have
identified two possible observable consequences, which may give birth
to an emergent subject of `D-foam phenomenology'.

One possible consequence is a linear energy-dependence of the velocity
of light~\cite{aemn,nature,Farakos} due to the interactions of photons with D-particle defects, which
would also depend linearly on the space-time density of these defects~\cite{emnnewuncert,mavro_review2009}.
Another possible consequence is a contribution to the vacuum energy
density (dark energy) that depends on the velocities of the D-branes and,
again, the density of D-particle defects~\cite{emninfl}. Therefore, between them,
measurements of the dark energy and of the velocities of energetic photons
could in principle constrain the density of D-particle defects and the
velocities of D-branes.

\begin{figure}[ht]
\centering
\includegraphics[width=7.5cm,angle=-90]{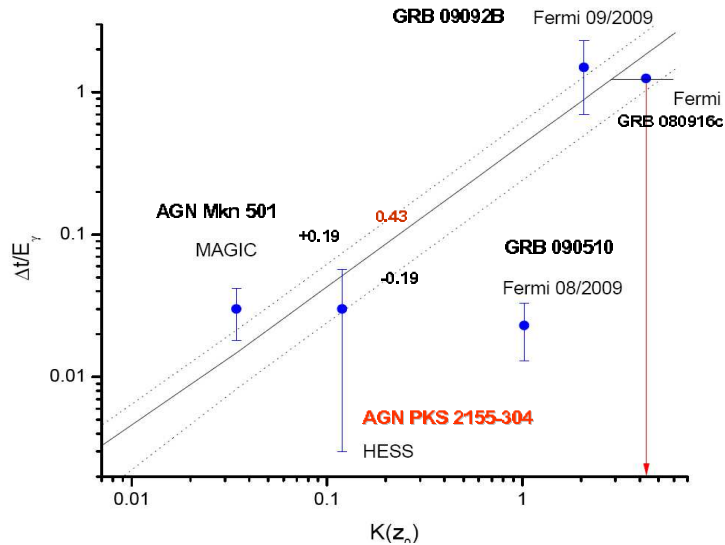}
\caption{\it Comparison of data on delays
$\Delta t$ in the the arrival times of energetic gamma rays from various astrophysical sources with
models in which the velocity of light is reduced by an amount linear in the photon
energy. The graph plots on a logarithmic scale the quantity $\Delta t/E$ and a function of the
redshift, $K(z)$, which is essentially the distance of the source from the observation point.
The data include two AGNs, Mkn 501~\protect\cite{MAGIC2} and PKS
2155-304~\protect\cite{hessnew}, and three GRBs observed by the Fermi satellite,
090510~\cite{grb090510}, 09092B~\cite{grb09092B} and 080916c~\cite{grbglast}.}%
\label{fig:data}%
\end{figure}

The experimental value of the present dark energy density $\Lambda$ is a fraction
$\Omega_\Lambda = 0.73 (3)$ of the critical density $1.5368 (11) \times 10^{-5} h^2$GeV/cm$^3$,
where $h = 0.73 (3)$, and the matter density fraction $\Omega_M = 0.27 (3)$.
The available cosmological data are consistent with the
dark energy density being constant, but some non-zero redshift dependence of
$\Lambda$ cannot be excluded, and would be an interesting observable in the
context of D-foam phenomenology, as we discuss later. The observational status of
a possible linear energy dependence of the velocity of light is less clear.
As shown in Fig.~\ref{fig:data}, observations of high-energy emissions from
AGN Mkn 501~\cite{MAGIC2} and PKS 2155-304~\cite{hessnew} are compatible with photon
velocities
\begin{equation}
v \; = \; c \times \left( 1 - \frac{E}{M_{QG}} \right) ,
\label{linear}
\end{equation}
where $\Delta t/E_\gamma = 0.43 (0.19) \times K(z)$~s/GeV,
corresponding to $M_{QG} =(0.98^{+0.77}_{-0.30}) \times 10^{18}$~GeV~\cite{emnnewuncert2}. 
This range is also
compatible with Fermi satellite observations of GRB 09092B~\cite{grb09092B}, 
from which at least one high-energy
photon arrived significantly later than those of low energies, and of GRB 080916c~\cite{grbglast}.
On the other hand, as also seen in Fig.~\ref{fig:data}, Fermi observations of GRB 090510~\cite{grb090510}
seem to allow only much smaller values of the retardation $\Delta t$,
and hence only values of $M_{QG} > M_P = 1.22 \times 10^{19}$~GeV.
However, these data probe different redshift ranges.

In this first combined exploration of D-foam phenomenology, we start by reviewing the
general connection between dark energy and a vacuum refractive index
in the general framework of D-branes
moving through a gas of D-particle defects in a 10-dimensional space. As we
discuss, there are various contributions to the dark energy that depend in
general on the density of defects and on the relative velocity of the D-branes.
On the other hand, the magnitude of the vacuum refractive index (\ref{linear}) is
proportional to the density of D-particle defects. We then
discuss the ranges of D-brane velocity and D-particle density that are
compatible with the measurements of $\Lambda$ and delays in the arrivals
of photons from AGN Mkn 501 and PKS 2155-304.
As seen in Fig.~\ref{fig:data}, these AGNs are both at relatively low redshift $z$,
where the experimental measurement of $\Lambda$ has been made, so the
same D-particle density is relevant to the two measurements. However,
it is not yet clear whether this value of $\Lambda$, and hence the same
D-particle density, also applied when $z \sim 1$.
If the density of D-particles was suppressed when $z \sim 1$ - a {\it D-void} - this could
explain~\cite{mavro_review2009} the much weaker energy dependence of the velocity of light allowed by
the Fermi observations of GRB 090510, which has a redshift of $0.903 (3)$,
as also seen in Fig.~\ref{fig:data}. In this case, the value of $\Lambda$
should also have varied at that epoch - a {\it clear experimental prediction of
this interpretation of the data}.
On the other hand, only a very abrupt resurgence of the D-particle density could
explain the retardation seen by Fermi in observations of GRB 09092b.
Alternatively, this retardation could be due to source effects - which should in any case also be
allowed for when analyzing the retardations seen in emissions from other sources.

\section{A D-Brane Model of Space-Time Foam and Cosmology}

\begin{figure}[ht]
\centering
\hspace{2cm}
\includegraphics[width=7.5cm]{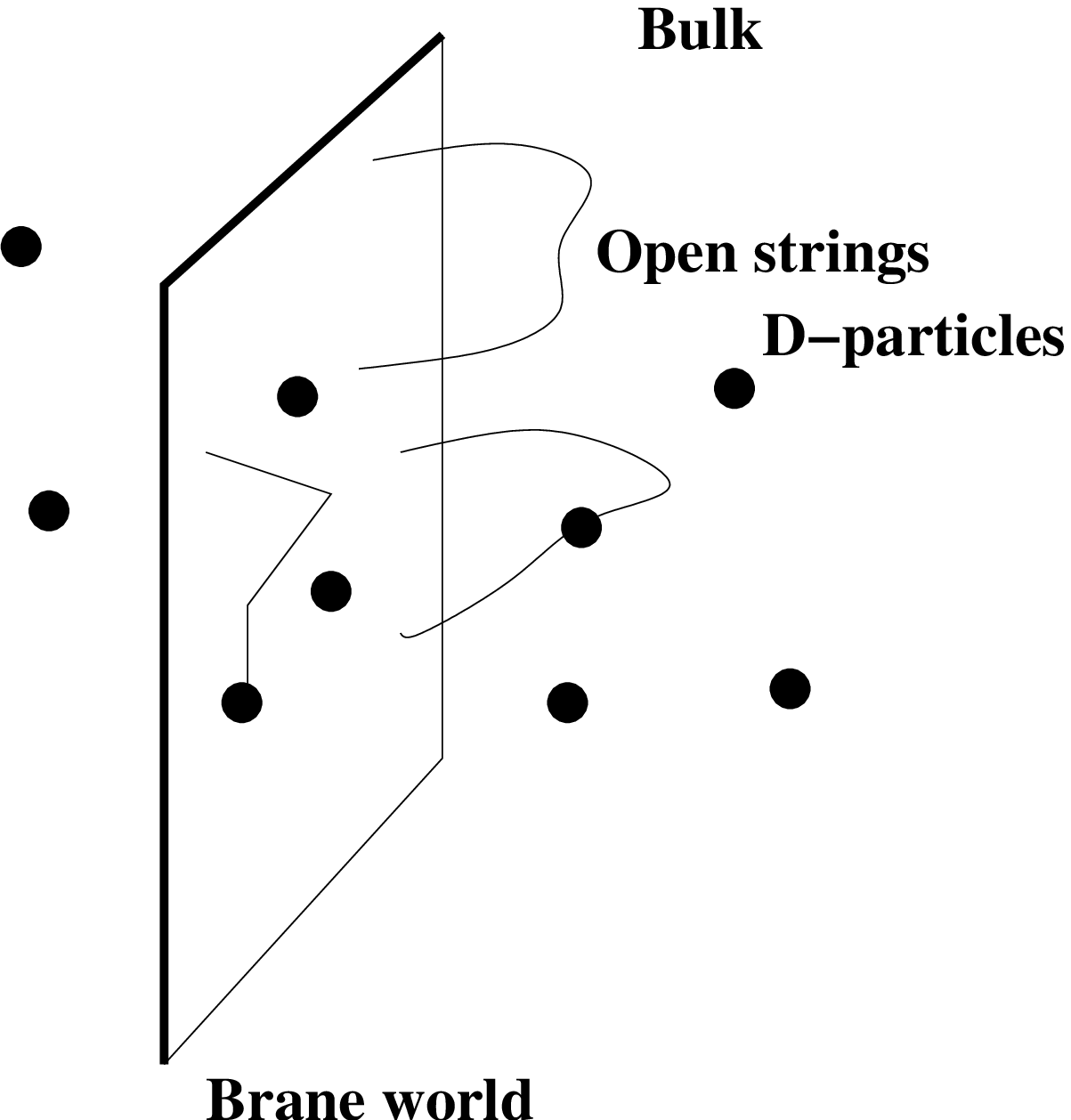} \hfill
\hspace{-5cm}
\includegraphics[width=7.5cm]{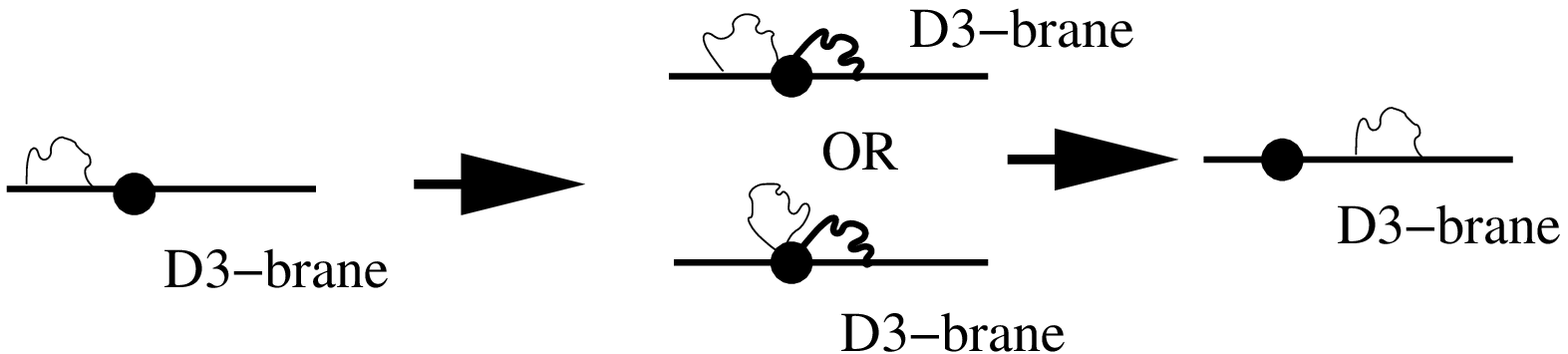}
\hspace{3cm}
\caption{\it Left: schematic
representation of a generic D-particle space-time foam model, in which matter particles
are treated as open strings propagating on a D3-brane, and the higher-dimensional
bulk space-time is punctured by D-particle defects. Right: details of the process
whereby an open string state propagating on the D3-brane is captured by a D-particle
defect, which then recoils.
This process involves an intermediate composite state that persists for a period
$\delta t \sim \sqrt{\alpha '} E$, where $E$ is the energy
of the incident string state, which distorts the surrounding
space time during the scattering, leading to an effective refractive index
but \emph{not} birefringence.}%
\label{fig:recoil}%
\end{figure}

As a concrete framework for D-foam phenomenology, we use the model illustrated in the left
panel of Fig.~\ref{fig:recoil}~\cite{Dfoam2,Dfoam}.
In this model, our Universe, perhaps after appropriate compactification, is represented  as a Dirichlet three-brane (D3-brane), propagating in a bulk
space-time punctured by D-particle defects~\footnote{Since an isolated D-particle cannot exist~\cite{polchinski},
because of gauge flux conservation, the presence of a D-brane is essential.}.
As the D3-brane world moves
through the bulk, the D-particles cross it. To an observer on the D3-brane the model looks
like `space-time foam' with defects `flashing' on and off as the D-particles cross it: this is the
structure we term `D-foam'. As shown in the left
panel of Fig.~\ref{fig:recoil}, matter particles are represented in this scenario by open strings
whose ends are attached to the D3-brane. They can interact with the D-particles
through splitting and capture of the strings by the D-particles, and subsequent
re-emission of the open string state, as illustrated in the right panel of Fig.~\ref{fig:recoil}. This
set-up for D-foam can be considered either in the context of type-IIA string
theory~\cite{emnnewuncert}, in which the D-particles are represented by
point-like D0-branes, or in the context of the phenomenologically more
realistic type-IIB strings~\cite{li}, in which case the D-particles
are modelled as D3-branes compactified around spatial three-cycles (in the simplest scenario),
since the theory admits no D0-branes. For the time being, we work in the type-IIA
framework,  returning later to the type-IIB version of D-foam phenomenology.

Within this approach, a photon propagating on the D3-brane (represented as an
open-string state) interacts with `flashing' D-particles at a rate proportional to their
local density $n_{D0}$. During each such interaction, a simple string
amplitude calculation or the application of the quantum-mechanical
uncertainty principle leads to the following order-of-magnitude estimate for
the time delay between absorption and re-emission of a photon of energy $E$:
\begin{equation}
\delta t_{D0} \; = \; C \sqrt{\alpha'} E ,
\label{delayD0}
\end{equation}
where $C$ is an unknown and model-dependent coefficient, and
$\alpha'$ is the Regge slope of the string, which is related to the
fundamental string length $\ell_*$ and mass $M_*$ by
$\sqrt{\alpha'} = \ell_* = 1/M_*$~\footnote{An estimate similar to (\ref{delayD0}) is also
obtained by considering the off-diagonal
entry in the effective metric experienced by the energetic photon due to the recoil of
the struck D-particle.}. This effect is independent of the photon polarization, and hence
does {\it not} yield birefringence.

When discussing D-foam phenomenology that involves
constraining the delay (\ref{delayD0}) accumulated over cosmological distances,
one must incorporate the effects of the redshift $z$ and the Hubble expansion rate $H(z)$.
Specifically, in a Roberston-Walker cosmology the delay due to any single scattering
event is affected by: (i) a time dilation factor~\cite{JP} $(1 + z)$ and (ii) the redshifting
of the photon energy~\cite{naturelater} that implies that the observed energy of a photon
with initial energy $E$ is reduced to $E_{\rm obs} = E_0/(1 + z)$. Thus, the observed
delay associated with (\ref{delayD0}) is:
\begin{equation}\label{obsdelay}
\delta t_{\rm obs} = (1 + z) \delta t_0 = (1 + z)^2 C \sqrt{\alpha '} E_{\rm obs} .
\end{equation}
For a line density of D-particles $n(z)$ at redshift $z$, we have $n(z) d\ell =  n(z) dt$ defects
per co-moving length, where $dt$ denotes the infinitesimal Robertson-Walker time interval
of a co-moving observer. Hence the total delay of an energetic photon in a co-moving
time interval $dt$ is given by $n(z) (1 + z)^2 C \sqrt{\alpha '} E_{\rm obs} \, dt $. The
time interval $dt$ is related to the Hubble rate $H(z)$ in the standard way:
$dt = - dz/[(1 + z) H(z)]$. Hence the total observed delay of a photon of observed
energy $E_{\rm obs}$ from a source at redshift $z$ till observation ($z=0$)
is~\cite{emnnewuncert,naturelater,mavro_review2009}:
\begin{equation}
\label{totaldelay}
\Delta t_{\rm obs} = \int_0^z dz C \frac{n(z)\, E_{\rm obs}}{M_s \,H_0} \,\frac{(1 + z)}{\sqrt{\Omega_M \, (1 + z)^3 + \Omega_\Lambda (z)}}~.
\end{equation}
In writing this expression, we assume Robertson-Walker cosmology, with $H_0$
denoting the present Hubble expansion rate, where $\Omega_M = 0.27$ is the present
fraction of the critical energy density in the form of non-relativistic matter, and
$\Omega_\Lambda (z)$ is the dark energy density. We assume that the
non-relativistic matter co-moves, with no creation or destruction, but leave open
the redshift dependence of the dark energy density, which we now
discuss.

\begin{figure}[ht]
\centering
\includegraphics[width=7.5cm]{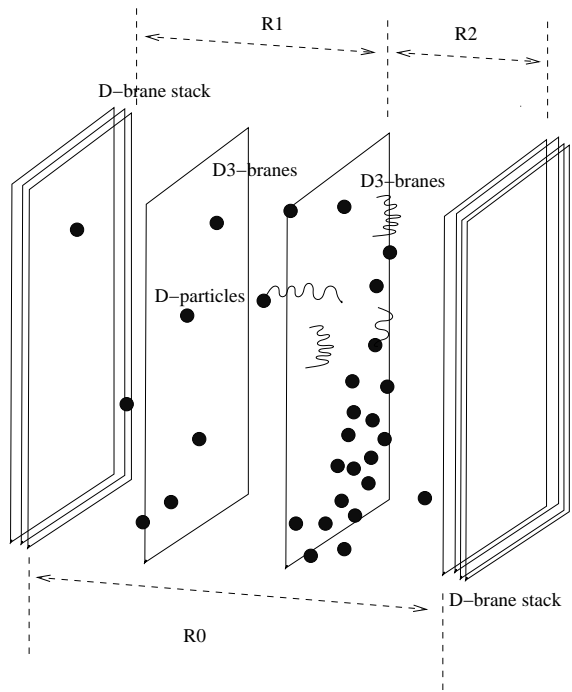} \caption{\it Schematic
representation of a D-particle space-time foam model with bulk density $n^\star (r) $ of D-particles
that may be inhomogeneous. The 10-dimensional space-time is bounded by two stacks
of D-branes, each accompanied by an orientifold.}
\label{fig:inhom}%
\end{figure}

The D3-branes appear in stacks, and supersymmetry dictates the number of
D3-branes in each stack, but does not restrict the number of D0-branes in the bulk.
For definiteness, we restrict our attention for now to the type-IIA model, in which the
bulk space is restricted to a finite range by two appropriate stacks of D8-branes, each
stack being supplemented by an appropriate orientifold eight-plane with specific
reflecting properties, so that the bulk space-time is effectively compactified to a finite
region, as illustrated in Fig.~\ref{fig:inhom}. We then postulate that two of the D8-branes
have been detached from their respective stacks, and are propagating in the bulk.
As discussed above, the bulk region is punctured by D0-branes (D-particles),
whose density may be inhomogeneous, as discussed below.
When there are no relative motions of the D3-branes or D-particles,
it was shown in~\cite{Dfoam} that the ground-state energy vanishes, as decreed by
the supersymmetries of the configuration. Thus, such static configurations constitute appropriate
ground states of string/brane theory. On the other hand, relative motions of the branes break
target-space supersymmetry explicitly, contributing to the dark energy density.
It is natural to assume that, during the current late era of the Universe, the D3-brane
representing our Universe is moving slowly and the configuration is evolving
adiabatically. One can calculate the vacuum energy induced on the brane world
in such an adiabatic situation by considering its interaction with the D-particles as well as the
other branes in the construction. This calculation was made in detail in~\cite{Dfoam},
where we refer the interested reader for further details. Here we mention only the results
relevant for the present discussion.

We concentrate first on D0-particle/D8-brane interactions in the type-IIA
model of~\cite{Dfoam}. During the late era of the
Universe when the approximation of adiabatic motion is valid, we use a weak-string-coupling
approximation $g_s \ll 1$. In such a case, the D-particle masses
$\sim M_s/g_s$ are large, i.e., these masses could be of the Planck size:
$M_s/g_s \sim M_P = 1.22 \cdot 10^{19}$~GeV or higher.
In the adiabatic approximation for the relative motion, these
interactions may be represented by a string stretched between the
D0-particle and the D8-brane, as shown in Fig.~\ref{fig:inhom}. The world-sheet amplitude of
such a string yields the appropriate potential energy between the D-particle and the D-brane,
which in turn determines the relevant contribution to the vacuum energy of the brane.
As is well known, parallel relative motion does not generate any potential, and the only
non-trivial contributions to the brane vacuum energy come from motion transverse
to the D-brane. Neglecting a velocity-independent term in the D0-particle/D8-brane
potential that is cancelled for a D8-brane in the presence of orientifold $O_8$
planes~\cite{polchinski}~\footnote{This cancellation is crucial for obtaining
an appropriate supersymmetric string ground state with zero ground-state energy.}, we find:
\begin{eqnarray}
\mathcal{V}^{short}_{D0-D8} & = & - \frac{\pi \alpha '}{12}\frac{v^2}{r^3}~~{\rm for} ~~
r \ll \sqrt{\alpha '}~,
\label{short}\\
\mathcal{V}^{long}_{D0-D8} & = & + \frac{r\, v^2}{8\,\pi \,\alpha '}~~~{\rm for} ~~
r \gg \sqrt{\alpha '}~.
\label{long}
\end{eqnarray}
where $v \ll 1$ is the relative velocity between the D-particle and the brane world, which is
assumed to be non-relativistic.
We note that the sign of the effective potential changes between short distances (\ref{short})
and long distances (\ref{long}). We also note that there is a minimum distance given by:
\begin{equation}
r_{\rm min} \simeq \sqrt{v \,\alpha '}~, \qquad v \ll 1~,
\label{newmin}
\end{equation}
which guarantees that (\ref{short}) is less than $r/\alpha'$, rendering
the effective low-energy field theory well-defined. Below this minimum distance,
the D0-particle/D8-brane string amplitude diverges
when expanded in powers of $(\alpha')^2 v^2 /r^4$. When they are separated from
a D-brane by a distance smaller than $r_{\rm min}$, D-particles should be considered as lying on
the D-brane world, and two D-branes separated by less than $r_{\rm min}$
should be considered as coincident.

We now consider a configuration with a moving D8-brane located at distances $R_{i}(t)$
from the orientifold end-planes, where $ R_1(t) + R_2(t) = R_0$ the fixed extent of the ninth bulk dimension, and the 9-density of the D-particles in the bulk is  denoted by $n^\star (r) $: see
Fig.~\ref{fig:inhom}. The total D8-vacuum-energy density $\rho^8$ - {\it D-energy} - due to the relative motions
is~\cite{Dfoam}:
\begin{equation}
  \mathcal{\rho}^{D8-D0}_{\rm total} = -  \int_{r_{\rm min}}^{\ell_s}  n^\star (r) \, \frac{\pi \alpha '}{12}\frac{v^2}{r^3}\, dr - \int_{-r_{\rm min}}^{-\ell_s}  n^\star (r) \, \frac{\pi \alpha '}{12}\frac{v^2}{r^3}\, dr
   +   \int_{-\ell_s} ^{-R_{1}(t)} \,n^\star (r) \frac{r\, v^2}{8\,\pi \,\alpha '}\, dr
+ \int_{\ell_s} ^{R_{2}(t)} \,n^\star (r) \frac{r\, v^2}{8\,\pi \,\alpha '}\, dr + \rho_0
\label{total}
\end{equation}
where the origin of the $r$ coordinate is placed on the 8-brane world and $\rho_0$
combines the contributions to the vacuum energy density from inside the band
$-r_{\rm min} \le r \le r_{\rm min}$, which include the brane tension.
When the D8-brane is moving in a uniform bulk distribution of D-particles,
we may set $n^\star(r) = n_0$, a constant, and the
D-energy density $\mathcal{\rho}^{D8-D0}_{\rm total}$ on the D8-brane is also
(approximately) constant for a long period of time:
\begin{equation}
\mathcal{\rho}^{D8-D0}_{\rm total} =  -n_0 \frac{\pi }{12} v(1 - v) + n_0 v^2 \frac{1}{16\pi \alpha '} (R_1(t)^2 + R_2(t)^2 - 2 \alpha ')
+ \rho_0~.
\label{total2}
\end{equation}
Because of the adiabatic motion of the D-brane, the time dependence of $R_i(t)$ is
weak, so that there is only a weak time dependence of the D-brane vacuum energy density:
it is positive if $\rho_0 > 0$, which can be arranged by considering branes with positive tension.

However, one can also consider the possibility that the D-particle density $n^\star (r)$
is inhomogeneous, perhaps because of some prior catastrophic cosmic collision,
or some subsequent disturbance. If there is a region depleted by D-particles - a {\it D-void} -
the relative importance of the terms in (\ref{total2}) may be changed.
In such a case, the first term on the right-hand side of (\ref{total2}) may
become significantly  smaller than the term proportional to $R_i^2(t)/\alpha'$.
As an illustration, consider for simplicity and concreteness a situation in which there
are different densities of D-particles close to the D8-brane ($n_{\rm local}$) and at long distances to
the left and right ($n_{\rm left}, n_{\rm right}$). In this case, one obtains  from (\ref{total}):
 \begin{equation}
\mathcal{\rho}^{D8-D0}_{\rm total} \simeq  -n_{\rm local} \frac{\pi }{12} v(1 - v) + n_{\rm left} v^2 \frac{1}{16 \pi \alpha '} \left(R_1(t)^2 - \alpha '\right)+  n_{\rm right} v^2 \frac{1}{16 \pi \alpha '} \left( R_2(t)^2 - \alpha '\right)
+ \rho_0~.
\label{total3}
\end{equation}
for the induced energy density on the D8-brane. The first term can be significantly smaller in
magnitude than the corresponding term in the uniform case, if the local density of D-particles
is suppressed. Overall, the D-particle-induced energy density on the D-brane world
{\it increases} as the brane enters a region where the D-particle density is {\it depleted}.
This could cause the onset of an accelerating phase in the expansion of the Universe. It is
intriguing that the redshift of GRB 090510 is in the ballpark of the redshift range where the
expansion of the Universe apparently made a transition from deceleration to acceleration~\cite{decel}.

The result (\ref{total3}) was derived in  an oversimplified case, where the possible
effects of other branes and orientifolds were not taken into account. However, as we now show,
the ideas emerging from this simple example persist in more realistic structures.
As argued in~\cite{Dfoam}, the presence of orientifolds, whose reflecting properties cause a
D-brane on one side of the orientifold plane to interact non-trivially with its image on the
other side, and the appropriate stacks of D8-branes
are such that the \emph{long-range contributions} of the D-particles to the D8-brane
energy density \emph{vanish}. What remain are the short-range D0-D8 brane
contributions and the contributions from the other D8-branes and O8 orientifold planes.
The latter are proportional to the fourth power of the relative velocity of the moving D8-brane world:
\begin{equation}
\mathcal{V}_{\rm{long} \,, \,D8-D8, , D8-O8} = V_8\frac{\left(aR_0 - b R_2(t)\right)v^4}{2^{13}\pi^9 {\alpha '}^5}~,
 \label{vlongo8d8}
 \end{equation}
where $V_8$ is the eight-brane volume, $R_0$ is the size of the orientifold-compactified
ninth dimension in the arrangement shown in Fig.~\ref{fig:inhom}, and the numerical coefficients in
(\ref{vlongo8d8}) result from the relevant factors in the appropriate amplitudes of strings in
a nine-dimensional space-time.
The constants $a > 0$ and $b >0$ depend on the number of moving D8-branes in the
configuration shown in Fig.~\ref{fig:inhom}.
If there are just two moving D8-branes that have collided in the past, as in the model we consider
here, then $a=30$ and $b = 64$~\cite{Dfoam}.
The potential (\ref{vlongo8d8}) is positive during late eras of the Universe
as long as $R_2(t) < 15R_0/32$. One must add to (\ref{vlongo8d8}) the negative
contributions due to the D-particles near the D8-brane world, so the total energy
density on the 8-brane world becomes:
\begin{equation}
\mathcal{\rho}_{8} \equiv \frac{\left(aR_0 - b R_2(t)\right)v^4}{2^{13}\pi^9 {\alpha '}^5} -n_{\rm short} \frac{\pi }{12} v(1 - v)~.
 \label{vtotal2}
 \end{equation}
As above, $n_{\rm short}$ denotes the nine-dimensional bulk density of D-particles near
the (moving) brane world.

As in the previous oversimplified example, the transition of the D8-brane world from a region
densely populated with D-particles to a depleted D-void causes these negative contributions
to the total energy density to diminish, leading potentially to an
acceleration of the expansion of the Universe. The latter lasts as long as the energy density
remains positive and overcomes matter contributions. In the particular example shown in
Fig.~\ref{fig:inhom}, $R_2(t)$ diminishes as time elapses and the D8-brane moves towards
the right-hand stack of D-branes, so the net long-distance
contribution to the energy density (\ref{total2}) (the first term) increases, tending further to
increase the acceleration.

It is the linear density of the D-particle defects $n(z)$ encountered by a propagating photon
that determines the amount of refraction. The density of D-particles crossing the D-brane
world cannot be determined from first principles,
and so may be regarded as a parameter in phenomenological models.
The flux of D-particles is proportional also to the velocity $v$ of the D8-brane in the bulk,
if the relative motion of the population of D-particles is ignored.

\section{Towards D-Foam Phenomenology}

In order to make some phenomenological headway,
we adopt some simplifying assumptions. For example, we may assume that between
a redshift $z < 1$ and today ($z=0$), the energy density (\ref{vtotal2}) has
remained approximately constant, as suggested by the available cosmological data.
this assumption corresponds to~\footnote{We recall that, for the case of two D8 branes
moving in the bulk in the model of~\cite{Dfoam}, we have $b=64 = 2^6$.}:
\begin{equation}\label{zero}
0 \simeq \frac{d\mathcal{\rho}_8}{d t } = H(z) (1 + z) \frac{d\mathcal{\rho}_8}{d z}  =
\frac{v^5}{2^{7}\pi^9 {\alpha '}^5} -\frac{d n_{\rm short}}{d t} \frac{\pi }{12} v(1 - v)~.
\end{equation}
where $t$ denotes the Robertson-Walker time on the brane world, for which
 $d/dt = -H(z) (1 + z) d/d z$, where $z$ is the redshift and $H(z)$ the Hubble parameter of the Universe,
and we take into account the fact that $dR_2(t)/dt = -v$ with $v > 0$,
due to the motion of the brane world towards the right-hand stack of branes in the model
illustrated in Fig.~\ref{fig:inhom}. Using $a(t) = a_0/(1 + z)$ for the cosmic scale factor,
equation (\ref{zero}) yields:
\begin{equation}\label{nchange}
\frac{v^5}{c^4\,2^{7}\pi^9 {\alpha '}^5} = -H(z) (1 + z) \frac{d n_{\rm short}}{d z} \frac{\pi }{12} \frac{v}{c}(1 - \frac{v}{c})~,
\end{equation}
where we have incorporated explicitly the speed of light \emph{in vacuo}, $c$.
Equation (\ref{nchange}) may be then integrated to yield:
\begin{equation}
\label{nofz}
n_{\rm short} (z) = n_{\rm short}(0) - \frac{12}{2^{7}\pi^{10}{\alpha '}^5} \frac{1}{\ell_s^9}\,\frac{c}{\ell_s}\, \frac{(v/c)^4}{(1 - \frac{v}{c})}\int_0^z \frac{d z'}{H(z') \, (1 + z')} \; \; {\rm where} \; \;  \ell_s \equiv \sqrt{\alpha '}~.
\end{equation}
In order to match the photon delay data of Fig.~\ref{fig:data} with the D-foam model,
we need a reduction of the \emph{linear} density of defects encountered by the photon by
about two orders of magnitude in the region $0.2 < z < 1$,
whereas for $z < 0.2 $ there must be, on average, one D-particle defect per unit string
length $\ell_s$. We therefore assume that our D-brane encountered a D-void when $0.2 < z < 1$,
in which there was a significant reduction in the bulk nine-dimensional density.

Ignoring the details of compactification, and assuming that a standard $\Lambda$CDM
cosmological model is a good underlying description of four-dimensional physics for $z < 1$,
which is compatible with the above-assumed constancy of
$\mathcal{\rho}_8$, we may write $H(z) = H_0 \sqrt{\Omega_\Lambda + \Omega_m (1 + z)^3}$,
where $H_0 \sim 2.5 \times 10^{-18} ~s^{-1}$ is the present-epoch Hubble rate.
Using (\ref{nofz}), and assuming $n_{\rm short}(0) = n^\star/\ell_s^9$, we obtain:
\begin{equation}
\label{nofz5}
n_{\rm short} (z) = \frac{n^\star}{\ell_s^9} \left( 1 - \frac{12}{2^{7}\pi^{10}}\,\frac{c H_0^{-1}}{n^\star \,\ell_s}\, \frac{(v/c)^4}{(1 - \frac{v}{c})}\int_0^z \frac{d z'}{(1 + z')\,\sqrt{\Omega_\Lambda + \Omega_m (1 + z')^3}}\right)~.
\end{equation}
It is clear that, in order to for $n_{\rm short}(z)$ to be reduced significantly, e.g.,
by two orders of magnitude, at the epoch $z \simeq 0.9$ of the GRB 090510~\cite{grb090510},
while keeping $n^\star ={\cal O}(1)$, one must consider
the magnitude of $v$ as well as the string scale $\ell_s$.
For instance, for string energy scales of the order of TeV, i.e., string time scales
$\ell_s/c = 10^{-27}~s$, one must consider a brane velocity $v \le \sqrt{10} \times 10^{-11} \, c$,
which is not implausible for a slowly moving D-brane at a late era of the Universe~\footnote{Much
smaller velocities are required for small string scales that are
comparable to the four-dimensional Planck length.}.
This is compatible with the constraint on $v$ obtained from inflation in~\cite{emninfl},
namely $v^2 \le 1.48 \times 10^{-5}\,g_s^{-1}$, where $g_s < 1$ for the
weak string coupling we assume here. On the other hand, if we assume a 9-volume
$V_9 = (K \ell_s)^9$: $K \sim 10^3$ and $\ell_s \sim 10^{-17}$/GeV, then the
brane velocity $v \le 10^{-4} \, c$.
In  our model, due to the friction induced on the
D-brane by the bulk D-particles, one would expect that the late-epoch brane velocity
should be much smaller than during the inflationary era immediately following
a D-brane collision.

\section{Towards D3-Foam Phenomenology}

The above discussion was in the context of type-IIA string theories, but may
easily be extended to compactified type-IIB theories of D-particle foam~\cite{li},
which may be of interest for low-energy Standard Model phenomenology.
In this case, as suggested in~\cite{li}, one may construct `effective' D-particle
defects by compactifying D3-branes on three-cycles. In the foam model
considered in~\cite{li}, the r\^ole of the brane Universe was played by D7-branes
compactified appropriately on four-cycles. The foam was provided by
compactified D3-brane `D-particles', in such a way that there is on average
a D-particle in each three-dimensional volume, $V_{A3}$, in the large
Minkowski space dimensions of the D7-brane. We recall that, although in the
conformal field theory description a D-brane is an object with a well-defined position,
in the full string field theory appropriate for incorporating quantum fluctuations in its position,
a D-brane is regarded as an object with thickness of the order of the string
scale. In particular, as discussed in~\cite{li}, an analysis of
the tachyonic lump solution in the string field theory,
which may be considered as a D-brane~\cite{Moeller:2000jy},
suggests that the widths of the D-brane in the transverse
dimensions are about $1.55\ell_s$.
In the setting of~\cite{li}, there are open strings between the D7-branes and D3-branes
that satisfy Neumann (N) and Dirichlet (D) boundary
conditions, respectively, which we call ND particles. Their gauge couplings with the
gauge fields on the D7-branes are
\begin{equation}
\frac{1}{g_{37}^2} = \frac{V}{g_7^2}~,~\,
 \label{coupl}
\end{equation}
where $g_7$ is the gauge coupling on the D7-branes, and
$V$ denotes the volume of the extra four spatial dimensions of the D7 branes
transverse to the D3-branes~\cite{giveon}. If there were no D-particle foam,
the Minkowski space dimensions would be non-compact, in which case
$V$ in (\ref{coupl}) would approach infinity and $g_{37} \to 0$.
In such a case, Standard Model particles would have no
interactions with the ND-particles on the compactified D3-brane (D-particle').
Thus, our Ansatz for the gauge couplings between
the gauge fields on the D7-branes and the ND-particles is~\cite{li}
\begin{equation}
\frac{1}{g_{37}^2} = \frac{V_{A3} R'}{(1.55\ell_s)^4}
\frac{\ell_s^4}{g_7^2} = \frac{V_{A3} R'}{(1.55)^4} \frac{1}{g_{7}^2}~.~\,
 \label{coupl-N}
\end{equation}
where $R'$ is the
radius of the fourth space dimension transverse to the D3-branes.
We remind the reader that the D7-brane gauge couplings are such that
$g_7^2 \propto g_s$, in accordance with general properties of D-branes
and their equivalence at low energies to gauge theories~\cite{giveon}.

The amplitudes describing the splitting and capture of a photon state by a
D-particle may be calculated in this picture using the standard techniques
outlined in~\cite{benakli}, with the result~\cite{li} that they are effectively
proportional to the coupling $g_{37}^2$ in (\ref{coupl-N}).
The amplitudes depend on kinematical invariants expressible in terms of the
Mandelstam variables: $s=-(k_1 + k_2)^2$, $t=-(k_2 + k_3)^2$ and $u=-(k_1 + k_3)^2$,
which satisfy $s + t + u =0$ for massless particles.
The ordered four-point amplitude $\mathcal{A}(1,2,3,4)$,
describing \emph{partly} the splitting and capture of a photon state by the D-particle, is given by
\begin{eqnarray}
\mathcal{A} (1_{j_1 I_1},2_{j_2 I_2},3_{j_3 I_3},4_{j_4 I_4}) = &&
- { g_{37}^2} l_s^2 \int_0^1 dx \, \, x^{-1 -s\, l_s^2}\, \, \,
(1-x)^{-1 -t\, l_s^2} \, \, \,  \frac {1}{ [F (x)]^2 } \, \times  \nonumber \\
 &&  \left[   {\bar u}^{(1)} \gamma_{\mu} u^{(2)}
{\bar u}^{(4)} \gamma^{\mu} u^{(3)} (1-x) + {\bar u}^{(1)} \gamma_{\mu}
u^{(4)} {\bar u}^{(2)} \gamma^{\mu} u^{(3)}  x \right ] \,  \nonumber \\
&&  \times \{ \eta
\delta_{I_1,{\bar I_2}} \delta_{I_3,{\bar I_4}}
\delta_{{\bar j_1}, j_4} \delta_{j_2,{\bar j_3}}
\sum_{m\in {\bf Z}}  \, \,
e^{ - {\pi} {\tau}\,
   m ^2 \, \ell_s^2 /R^{\prime 2}   }
\nonumber \\
&& +  \delta_{j_1,{\bar j_2}}
\delta_{j_3,{\bar j_4}}
\delta_{{\bar I_1}, I_4} \delta_{I_2,{\bar I_3}}
\sum_{n\in {\bf Z}}  e^{-  {\pi \tau}   n^2
\,  R^{\prime 2} / \ell_s^{2} } \}~,~\,
\label{4ampl}
\end{eqnarray}
where $F(x)\equiv F(1/2; 1/2; 1; x)$ is the hypergeometric function,
$\tau (x) = F(1-x)/F(x)$,
$j_i$ and $I_i$ with $i=1, ~2, ~3, ~4$
are indices on the D7-branes and D3-branes, respectively, and $\eta=(1.55\ell_s)^4/(V_{A3} R')$
in the notation of \cite{benakli}, $u$ is a fermion polarization spinor,
and the dependence on the appropriate Chan-Paton factors has been made
explicit~\footnote{This is only part of the process. The initial splitting of a photon into two open
string states and their subsequent re-joining to form the re-emitted photon both depend on the
couplings $g_{37}^2$, so there are extra factors proportional to $\eta$ in a complete treatment,
which we do not discuss here.}.
In the above we considered for concreteness the case where the photon state splits into
two fermion excitations, represented by open strings. The results are qualitatively
identical in the case of boson excitations. In fact, the spin of the incident open strings does not
affect the delays or the relevant discussion on the qualitative features of the effective
number of defects interacting with the photon. Technically, it should be noted that the
novelty of our results above, as compared with those of~\cite{benakli},
lies in the specific compactification procedure adopted, and the existence of a
uniformly-distributed population of D-particles (foam), leading to (\ref{coupl-N}).
It is this feature that leads to the replacement of the simple string coupling $g_s$ in the Veneziano amplitude by the effective D3-D7 effective coupling $g_{37}^2$.

In this approach, time delays for photons arise~\cite{sussk1} from  the
amplitude $\mathcal{A}(1,2,3,4)$ by considering
backward scattering $u=0$ and looking at the corresponding pole structure.
The details of the discussion
are presented in~\cite{li} and are not repeated here. The photon delays are
proportional to the incident energy, and are suppressed by a single power of the
string scale, as discussed above.
Due to the $\eta$-dependent terms in (\ref{4ampl}),
charged-particle excitations in the low-energy limit, such as electrons, have
interactions with the D-particle foam that are non-zero in type-IIB models,
in contrast to the type-IIA models discussed previously.
However, these interactions are suppressed by further factors of $\eta$ compared to photons.
The stringent constraints from the Crab-Nebula synchrotron-radiation observations~\cite{crab}
are satisfied for $\eta ={\cal O}(10^{-6})$, which is naturally obtained in such
models~\cite{li}~\footnote{As already mentioned, the ordered amplitude (\ref{4ampl})
describes only part of the process, and there are extra factors of $\eta$ coming from the initial
splitting of the photon into two open string states. When these are taken into account, the
constraints from Crab Nebula~\cite{crab} can be satisfied for much larger values of $\eta$.
We postpone a more detailed discussion of D3-foam phenomenology for future work.}.

The cosmological scenario of a D-brane moving in a bulk space punctured by
D-particles can be extended to this type-IIB construction.
If there was a depletion of D-particle defects
in a certain range of redshifts in the past, e.g., when $z \sim 0.9$,
the volume $V_{A3}$ in (\ref{coupl-N}) would have
been larger than the corresponding volume at $z \ll 1$,
and the corresponding coupling reduced, and hence also the corresponding cross section
$\sigma$ describing the probability of interaction of a photon with the D-particle in the foam.

The \emph{effective} linear density of defects $n(z)$ appearing in (\ref{totaldelay})
can then be related to the number density $\tilde{n}^{(3)}$ per unit three-volume on the
compactified D7-brane via $ n(z) \sim \sigma \tilde{n}^{(3)}$. The quantity $\tilde{n}^{(3)}$ is,
in turn, directly related to the bulk density of D-particles in this model, where there is no
capture of defects by the moving D-brane, as a result of the repulsive forces between them.
The cross section $\sigma$ is proportional to the square of the string amplitude (\ref{4ampl})
describing the capture of an open photon string state by the D-particle in the model,
and hence is proportional to $g_{37}^4$, where $g_{37}$ is given in (\ref{coupl-N}).
If recoil of the D-particle is included,
the amplitude  is proportional to the effective string coupling: $g_s^{\rm eff} \propto g_{37}^2$,
where $g^{\rm eff} = g_s (1 - |\vec u|^2)^{1/2}$, with $\vec u \ll 1$  the recoil velocity
of the D-particle~\cite{emnnewuncert,li}.
One may normalize the string couplings in such a way that $n(z=0)={\cal O}(1)$,
in which case $n(z\simeq 0.9) \ll 1$ by about two orders of magnitude.
However, in general the induced gauge string couplings depend crucially on the
details of the compactification as well as the Standard Model phenomenology on the
D-brane world. Hence the precise magnitude of the time delays is model-dependent,
and can differ between models. We defer a detailed discussion of such issues to a future work,
limiting ourselves here to noting that photon delay experiments place crucial
constraints on such diverse matters as the dark sector of a string Universe and its
compactification details.

\section{Conclusions}

The above examples are simplified cases capable of an analytic treatment,
in which the bulk density of D-particles may be determined phenomenologically by
fitting the photon delay data of Fig.~\ref{fig:data}, and those to be obtained by future
measurements. In a more complete theory, the late-era bulk distribution of defects
would be determined by the detailed dynamics and the subsequent evolution of
disturbances in the initial population of D-particles after the brane collision.
Therefore, one needs detailed numerical simulations of the gas of D-particles in such
colliding D-brane models, in order to determine the relative time scales involved in the
various phases of the D-brane Universe in such foamy situations. These are complicated by
the supersymmetric constructions and the orientifold planes involved, but may be
motivated by the important r\^oles that the D-particles can play in providing dark energy,
as we, as making a link to a seemingly unrelated phenomenon, namely a possible
refractive index for photons propagating {\it in vacuo}.

Already in this first discussion of D-foam phenomenology,
we have established that the available data are not incompatible with a refractive
index that varies {\it linearly} on the photon energy, with a coefficient that depends
on the density and coupling of the D-particle defects. These may depend on the
cosmological epoch, and their density may also influence the magnitude of the
cosmological dark energy density. We defer to later work a more detailed
fit to the photon delays and dark energy observed. {\it Une affaire \`a suivre...}

\section*{Acknowledgements}

The work of J.E. and N.E.M. is partially supported by the European Union
through the Marie Curie Research and Training Network \emph{UniverseNet}
(MRTN-2006-035863), and that of D.V.N. by
DOE grant DE-FG02-95ER40917.

\end{document}